\documentstyle[twocolumn,aps,prl,epsf]{revtex}

\begin{document}

\title{Monte Carlo procedure for protein folding in
lattice models. Conformational rigidity.}

\author{Olivier Collet}

\address{Laboratoire de Biophysique Mol\'eculaire, UMR CNRS 7565,
Facult\'{e} des Sciences, \\
Universit\'{e} Henri Poincar\'{e}-Nancy 1,
54506 Vandoeuvre-l\`{e}s-Nancy, France}
\address{
\centering{
\medskip\em
{}~\\
\begin{minipage}{14cm}
A rigourous Monte-Carlo method for protein folding simulation on 
lattice model is introduced.  We show that a parameter 
which can be seen as the rigidity of the conformations has to be
introduced in order to satisfy the detailed balance condition.
Its properties are discussed and its role during 
the folding process is elucidated.
This method is applied on small chains on two-dimensional lattice. 
A Bortz-Kalos-Lebowitz type algorithm which allows to study the
kinetic of the chains at very low temperature is implemented
in the presented method. We show that the coefficients of the
Arrhenius law are in good agreement with the value 
of the main potential barrier of the system.
{}~\\
{}~
\end{minipage}
}
}

\maketitle
\newpage

%\section{Introduction}
 
Proteins are heteropolymers that exbihit surprising 
thermodynamic and kinetic properties.
The first aspect is that the lowest free energy conformation of a protein 
is assumed to be the unique native structure and to be thermodynamically stable
\cite{Anfinsen1961}.
A major challenge in theoretical protein folding is to understand the 
second aspect or in other words, how does a 
protein find its native structure in biologically reasonnable 
times under physiological conditions \cite{Levinthal1968}. 
The lattice model is one class of models that is used to study 
theoretically the folding of protein \cite{Lau1989,Sali1994a,Sali1994b}
and Monte Carlo (MC) algorithms \cite{Metropolis1953} are widely
used to study dynamics \cite{Chan1993,Chan1994,Socci1994,Gutin1998}.

In this Letter, we show that the commonly used 
MC procedure converges poorly towards thermal equilibrium.
An attempt to refine the procedure has been recently proposed 
by Cieplak et al.  \cite{Hoang1998,Cieplak1999}, but
even if this procedure converges towards equilibrium,
the parameters of the Arrhenius law that they found disagree with the
value of the main potential barrier obtained independently 
by a study of the phase space of the systems.
We introduce, here, a more rigorous treatment of the dynamics. 
Our method fulfil the detailed balance condition, and, then,
converges, indeed, towards the thermal equilibrium.
For the first time, it also shows a good efficiency in the
calculation of kinetics parameters and the determination 
of the Arrhenius law.

%\section{Model}

The model used is a two-dimensional lattice polymer. The chains are
composed of $N$ monomers that are connected and constrained to be on a
square lattice and the chains are self avoiding walk.
The energy of a sequence in a given conformation $m$ is given by: 

\begin{equation}
E^{(m)}=\sum_{i > j+1} (B_{ij}+B_0) \Delta^{(m)}_{ij}
\end{equation}
where the function $\Delta^{(m)}_{ij}$ equals 1 if the $i^{th}$ and $j^{th}$
monomers interact i.e. if they are nearest neighbors on the lattice. The 
$B_{ij}'s$ are the contact energy values. They are chosen randomly in
a gaussian distribution centered on $0$, and 
$B_0$ is a negative parameter which favors the compact conformations
\cite{Shakhnovich1990a,Dinner1994}. 
The set of $B_{ij}$ gives a sequence of the chain. 

The sets of connections between conformations, used for the MC procedure,
are those used by Chan and Dill \cite{Chan1994} : 
the corner flip and the tail moves are referred to as the
move set 1 (MS1), the crankshaft move is referred to as the move set 2 (MS2) and
at each MC step, a move of MS1 is chosen with a probability $r$
and a move of MS2  is chosen with a probability $1-r$ \cite{Sali1994a}.

%\section{Theory}

Now, the problem is to find a correct algorithm of Metropolis \cite{Metropolis1953}
which
guarantees that the simulation converges towards thermal equilibrium
imposed by the condition of the detailed balance :
\begin{equation}
  P_{eq}^{(m)} W(m\rightarrow n) = P_{eq}^{(n)} W(n \rightarrow m)
\end{equation}
where $P_{eq}^{(m)} \propto \exp(-E^{(m)}/T)$ is the equilibrium probability 
of the conformation $m$, $T$ is the temperature,
and $W(m\rightarrow n)$ is the probability of transition from the state $m$
to the state $n$. Let us note :
\begin{equation}
\label{wmn}
  W(m\rightarrow n) =  
W^{(0)}(m\rightarrow n) \, \, a(m\rightarrow n)
\end{equation}
where $W^{(0)}(m\rightarrow n)$ is the a priori transition probability.
A convenient choice for the  acceptance ratio is~: 
\begin{equation}
a(m\rightarrow n) = \frac {1} {1 + \exp(\Delta E_n^m / T)}
\end{equation}
with $\Delta E_n^m = E^{(n)} - E^{(m)}$.
Let us note $N_m^{(1)}$ and $N_m^{(2)}$ the number of allowed 
transitions from $m$ to any conformation by performing a move of the MS1 or of 
the MS2 and $N_{max}^{(1)}=\max_m\{N_m^{(1)}\}$ and $N_{max}^{(2)}
=\max_m\{N_m^{(2)}\}$. One can easily see that
$N_{max}^{(1)} = N+2$  and $N_{max}^{(2)} = N-7$.
In order to have symmetric a priori transition probabilities~: 
$W^{(0)}(m\rightarrow n) = W^{(0)}(n \rightarrow m)$, one assumes
that the probability to attempt a move from conformation $m$ to conformation
$n$ related by a connection of respectively MS1 and MS2 during one MC step are then~:
\begin{equation}
\label{w01}
W^{(0)}_1 (m\rightarrow n) = \frac{r} {N_{max}^{(1)}}= \frac{r}{N+2}
\end{equation}
\begin{equation}
\label{w02}
W^{(0)}_2 (m\rightarrow n) = \frac{(1-r)} {N_{max}^{(2)}}=\frac{1-r}{N-7}
\end{equation}

%FIGURE 1
\begin{figure}
\epsfxsize=3.2in
\epsfysize=0.7in
\centerline{\epsffile{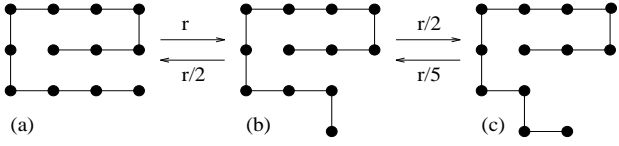}}
\vskip 0.5cm
\caption{A part of the connection graph of the 12 monomers chain.
The conformations (a), (b) and (c) are connected to  respectively one, two and five
neighbors by MS1.
In the classical MC procedure, the a priori transition probabilities
are not symmetric.
They depend on the number of neighbors.
With the proposed method, all these transitions are attempted
with the same a priori transition probability (not show).
}
\label{neigh}
\end{figure}
    
Then, the probability to attempt any move from the conformation $m$ using the
MS1 is $r N^{(1)}_m /(N+2)$ (and $(1-r) N^{(2)}_m /(N-7)$ using the MS2). 
And therefore, it appears a probability of null transition :
\begin{equation}
  w_m^{(0)} = 1 - \left( r \frac{N_m^{(1)}} {N+2}
  + (1-r)  \frac{N_m^{(2)}} {N-7} \right)
\end{equation}

In contrast with rigid rotation which can involve movements of a lot of monomers, 
the one and two monomers moves are local modifications. One assumes, then, that they 
have the same affinity. Then, it comes from equations \ref{w01} and \ref{w02} :
\begin{equation}
  r = \frac{N+2}{2N-5}
\end{equation}
In this particular case, the previous equations simplify  :
\begin{equation}
 W^{(0)}_1 (m\rightarrow n) = W^{(0)}_2 (m\rightarrow n) =
\frac{1}{2N-5}
\end{equation}
\begin{equation}
  w_m^{(0)} = 1 - \frac{N_m^{(1)} + N_m^{(2)}}{2N-5}
\end{equation}

In order to check the accuracy of the proposed procedure, we applied it on
12 monomers chains. These chains can adopt 15037 different self avoiding walk
conformations non equivalent by symmetry.  The following results are obtained
for the sequence A defined elsewhere \cite{Cieplak1998}.    
Such a short chain is used to check the method
because a convergence test can be applied to this chain in a reasonable
computational time.

For this chain, we performed 300 billions steps MC trajectories. 
A convergence factor $C(t) = \sqrt{\langle (P_{eq}^{(m)} - {\rm occ}^{(m)}(t))^2
\rangle }$ is computed 
each 100000 MC steps ;
$t$ is for number of the MC step and ${\rm occ}^{(m)}(t)=N^{(m)}(t)/t$ where
$N^{(m)}(t)$ is the number of steps corresponding to the occurences
of the conformation $m$.
The brackets denote the average over all the conformations. 
If a simulation checks well the detailed balance, the $C(t)$ quantities 
should tend towards 0 when $t \rightarrow \infty$. 

%FIGURE 1
\begin{figure}
\epsfxsize=3.2in
\epsfysize=3.2in
\centerline{\epsffile{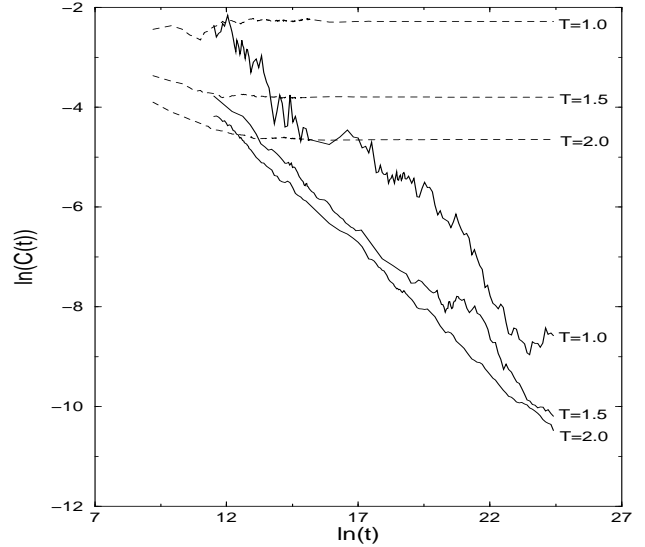}}
\caption{
    Log-Log plots of the convergence factor $C(t)$ versus the 
    number of MC step $t$ for different temperatures.  
    Dashed lines : for the commonly used method for which the 
    $W^{(0)}(m \rightarrow n)$
    prefactor and $w^{(0)}_m$ parameter are omitted
    in the MC procedure. 
    Solid lines : for the proposed method. 
}
\label{converg_A}
\end{figure}

Figure ~\ref{converg_A} shows clearly that the 
commonly used procedure present limits of convergence depending on the
temperature. On the other hand, the proposed method shows a power law of 
the convergence factor versus the MC steps. 
This result shows very well that the factor $w^{(0)}_m$
cannot be omitted in a lattice simulation for
protein folding.

\begin{figure}
\vskip -2.5cm
\epsfxsize=3.2in
\epsfysize=4.8in
\centerline{\epsffile{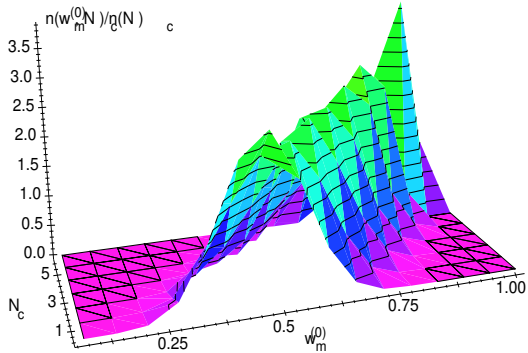}}
\vskip-4.5cm
\caption{
Distribution of the number of conformations as a function of 
the rigidity $w_m^{(0)}$ and the  number of intrachain contacts
$N_c$ divided by the number of conformations which have 
$N_c$ contacts.
One must note that the smaller the number of contacts of the
conformations of a subset the larger the normalisation factor. 
}
\label{distribution}
\end{figure}

In what follows, we focus on the properties of the $w^{(0)}_m$ factor. 
One must notice that this factor is only a topological factor and 
therefore is sequence independent.
If one looks now at the simulation only at the topological point of
view, by removing for a while the energetic contribution (let suppose
for a while that all conformations have the same energy), one sees
that, the larger the factor $w^{(0)}_m$, the longer the simulation 
stays in the $m$ conformation when it reaches it. 
One must note that it is not only unprobable to escape from the conformation
$m$ if $w^{(0)}_m$ is large, but it is also unprobable to reach it.
On the contrary, conformations with small values of $w^{(0)}_m$ are often
reach but the simulation doesn't stay in this conformation.
Then, the larger $w^{(0)}_m$, the more rigid the conformation $m$
and the smaller $w^{(0)}_m$, the more flexible the conformation $m$.
Therefore, let us call in what follows $w^{(0)}_m$ the rigidity of the
conformation $m$.
Fig.~\ref{distribution}(b) show how the $w^{(0)}_m$
prefactors are distributed for each subset of conformations with the same number of
contacts. 
No conformation have a value of $w^{(0)}_m$ equal to 1(fig.
 \ref{distribution}). 
This guarantees that, no conformation is totally rigid,
then each one is related to at least, another one. But, one must note that
this condiction is not enough strong to fulfil the ergodic hypothesis.

%FIGURE 2
It appears clearly that, the more compact conformations
present the larger values of the rigidity.
The more flexible are the more extended. 
Only one move of MS1 is allowed for the two more rigid conformations.
One can see that, there is no conformation which have a value of $w^{(0)}_m$
which tends towards 0. Hence no conformation is totally flexible.
This is a consequence of that no conformation present the maximum
number of neighbors with both the MS1 and the MS2. 

The native conformations of protein has not only very low energy but
also they are very compact. Hence, they have 
of large Boltzmann weights but also they are very rigid conformations.
Both effects favorise the stability of the native conformations, but
the folding dynamics is slowed down by the topology of the native structures.
The trap conformations are also very compact and are conformation of
local energy minima \cite{Cieplak1998}. 
Then to exit the trap valley the chain has to first escape from a 
stable and rigid conformation.

One computed many kinetic ways from the trap conformation to the native
structure of the sequence A. The trap conformation has been determined
by solving the master equation of the system using the way described 
by Cieplak et al \cite{Cieplak1998} for the particular choice of $r$ 
used in the present paper.
The conformation trap found here is the same that the conformation 
found by Cieplak et al.  and it is chosen as the first conformation
of the MC trajectories.
The kinetic ways exhibit all similar properties. 
The native and the trap conformations are
compact and then very rigid ($w^{(0)}_{natif} = w^{(0)}_{trap}= 0.894$)
and have low energies. 
%FIGURE 4
\begin{figure}
\epsfxsize=2.4in
\epsfysize=3.0in
\centerline{\epsffile{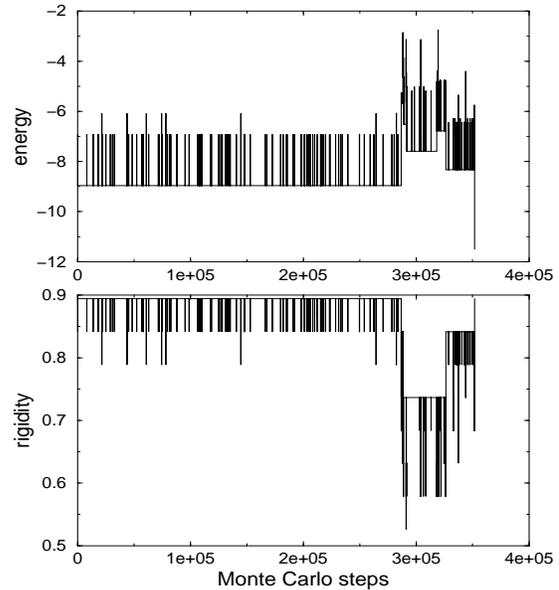}}
\vskip 0.5cm
\caption{
Energy (top) and rigidity (bottom) versus the MC steps
of a typical trajectory of folding simulation for the sequence A
at $T=0.4$.
}
\label{reaction_way}
\end{figure}
At low temperature the system spends a lot
of MC steps in the trap conformation. The system escapes uneasily 
from the trap
by passing through  transition states which exhibit common properties :
high energies, few intrachain contacts and then great flexibility. 
Therefore, even if  the transition states are energetically unfavorable, 
they are easily accessible at a topological point of view and the 
MC trajectories spend a very few steps in these conformations.

A major problem of the protein folding investigation is namely to
calculate kinetic properties at low temperature \cite{Cieplak1996,Gutin1998}, 
where the rejected move ratio of a MC procedure is 
very large. The efficiency of the procedure is increased at low temperature 
using a Bortz-Kalos-Lebowitz (BKL) type algorithm \cite{Bortz1975}.
The idea is the following : let us note, $w_m$ the probability not to 
accept a move from the conformation $m$ during one step~:
\begin{equation}
  w_m = 1 - \frac{1}{2N-5} \sum_{n \neq m} \frac{1}{1+\exp (
  \Delta E_{n}^m/T)}
\end{equation}
then, the probability not to accept a move from the conformation
$m$ during exactly $k$ steps is :
\begin{equation}
  P(k) = w_m^{k-1} (1-w_m)
\end{equation}
Then for each move, the number of MC steps $k$, during which the chain 
stays in the current conformation, say $m$, is chosen at random in
the density of probability $P(k)$ and a move chosen with the following
probability of transition : 
\begin{equation}
  t(m\rightarrow n) = {{\frac{1}{1+\exp (\Delta E_{n}^m/T)}}
  \over
  {\sum_{n' \neq m} \frac{1}{1+\exp (\Delta E_{n'}^m/T)}}}
\end{equation}
is always performed. This procedure permits to carry out 
MC simulations at very low temperature.
All the values of $w_m$ and $t(m\rightarrow n)$ are computed for each
temperature before performing the MC trajectories.

The folding times ($t_{fold}$) have been computed using the BKL type 
algorithm for low temperature. 
The folding time is the average over 500 trajectories
of the number of MC steps needed to reach the
conformation of lowest energy.
Three different simulations have been
carried out depending on the choice of the first conformation set ; 
the simulation "T" for which the trap conformation is 
chosen as the first conformation ; 
the simulation "E" for which the first conformation is an extended 
conformation chosen at random ; 
the simulation "R" for which the first 
conformation is chosen at random among all the conformational space.
The transition state of lowest energy between the trap and the 
native structure has been determined elsewhere for this 
sequence \cite{Cieplak1999} 
and the difference of energy between the trap and the transition state had been
computed and equals $\Delta E = 4.53$. The Monte-Carlo folding time
found by Cieplak et al. follows an Arrhenius law $t_{fold}(T) = A \exp (\delta E /T)$,
with $\delta E = 2.76$ which is in poor agreement with $\Delta E$.
For the three simulations, we also find Arrhenius laws
at very low temperature ($T=$ 0.24, 0.22, 0.20, 0.18). 
\vbox{
\begin{table}[t]
\begin{center}
\begin{tabular}{ c c c } 
      &  $\delta E$ &  A \\  \hline
 simulation "T"   &  4.51     &  33.25 \\
 simulation "E"   &  4.40     &  8.58  \\
 simulation "R"   &  4.34     &  12.55 \\
\end{tabular}
\caption{Value of the parameters $\delta E$ and $A$ of the Arrhenius laws 
$t_{fold}(T) = A \exp (\delta E /T)$ for the "T", "E" and "R"
simulations (see text)
}
\end{center}
\end{table}
}
\vskip -1cm
The results of $\delta E$, shown in table 1, are in very good agreement 
(less than $1 \%$ for the "T" simulation) with the value of 
$\Delta E$ and strongly support the proposed method for
the calculation of the parameters of the Arrhenius laws.

If a first conformation is chosen at random, it can fall in the trap
valley (TV), in the native conformation valley (NV) or in less important
valleys. At low temperature, whatever the set of first conformations, the
conformations which fall in TV govern the kinetics. Then, the dominant
term in the exponential function of the Arrhenius law tends always towards
$\Delta E$. The ratio of the $A$ coefficients gives the proportion 
of conformations which falls in TV :   
a random conformation has a probability equal
to $12.55 / 33.25 = 0.38$ to fall in TV and an extended 
conformation has a probability equal to 0.26 to be attracted by TV.
Then, these ratios give an insight of the attraction strenght of the basin of TV.

The results presented in this Letter show clearly that the proposed MC method
is well adapted to the study of the dynamics of protein folding.
It has been showed that not only the difference
of energy between the conformations has to be taken into account in
the MC simulations but also the rigidity of the conformations.
The method had been applied only on a short chain in order to check
its efficiency, but it is 
easily applicable to longer chain on two- or three-dimensional lattices and
moreover the BKL algorithm should permit to elucidate low temperature properties
of protein-like chains. 

{\bf Acknowledgments} to Aaron Dinner, Bertrand Berche, Christophe Chatelain, 
Trinh Xuan Hoang and Marek Cieplak
for helpful discussions.

%\bibliography{../../biblio}

\end{document}